# The Role of Stellar Mass and Star Formation in Shaping X-ray Emission of Radio-Loud and Radio-Quiet AGN

SH. M. Shehata[a], F. Shaban[b,*], R. M. Samir[a]

[a]*National Research Institute of Astronomy and Geophysics (NRIAG) Astronomy Department Cairo Egypt*
[b]*Astronomy Space Science and Meteorology Department Faculty of Science Cairo University Giza Egypt.*

**Abstract**

Galaxy evolution and extragalactic astronomy research depend on an understanding of the interactions between active galactic nuclei (AGN) and their host galaxies. We investigate the relationships between X-ray luminosity ($L_X$), star formation rate (SFR), and stellar mass (M∗) in distinct samples of radio-loud (RL) and radio-quiet (RQ) AGN. Using data from 4XMM-DR11, SDSS-DR16, and the DESI AGN Host Galaxies VAC, we examine how these key properties correlate within each AGN population. Our analysis reveals different behaviors: RL-AGN show a strong, statistically significant $L_X$–SFR correlation but no significant link with M∗, suggesting that accretion and star formation are coupled, possibly independent of host mass. In contrast, RQ-AGN display moderate, significant positive correlations across all parameters, consistent with joint growth driven by a shared cold gas supply. These results suggest that radio-loud AGN might slow down star formation in their galaxies, while radio-quiet AGN seem to grow together with it.

## 1. Introduction

Active galactic nuclei (AGN) are considered to be one of the most luminous and dynamic objects in the universe, as they are powered by the accumulation of matter onto supermassive black holes. Their emissions cover a wide range of the electromagnetic spectrum, X-ray emission in AGNs arises from the corona near the accretion disk and where high-energy processes produce X-rays. Investigating the correlation between star formation rate (SFR) and stellar mass of galaxies that host AGN is essential to gain awareness into galaxy growth (Rosario et al., 2013; Bernhard et al., 2019; Koutoulidis et al., 2022; Mountrichas et al., 2024). The connection between stellar mass and the galaxys´ evolutionary status is complex and significant. Research has shown a strong correlation between SFR and stellar mass, with larger galaxies typically exhibiting more active star formation. This highlights the importance of AGN in controlling star formation, potentially through processes like AGN-induced outflows that can either suppress or stimulate star formation depending on the circumstances.

In a similar vein, a correlation is observed between the SFR and X-ray luminosity in galaxies hosting AGN. Research has shown that X-ray luminosity, which is a reliable indicator of the accretion activity surrounding supermassive black holes, tends to increase in tandem with the SFR (Mullaney et al., 2012; Rosario et al., 2013; Heinis et al., 2016; Masoura et al., 2018; Stemo et al., 2020; Torbaniuk et al., 2021, 2024). This connection implies a mutually beneficial relationship where active star-forming regions may contribute to the AGN's fuel source, while the AGN, in turn, impacts star formation through feedback mechanisms. Thus, the categorization of AGN into radio-loud and radio-quiet groups indicates fundamental differences in the physical characteristics and emission mechanisms. This classification allows us to explore the distinct X-ray emission mechanisms in these two types of AGN. Radio-loud AGN often show enhanced X-ray luminosities due to contributions from relativistic jets, whereas radio-quiet AGN primarily emit X-rays from the accretion disk and corona (Brinkmann et al., 2000). In contrast, AGN that are classified as radio quiet showed evidence suggesting a positive relationship between X-ray luminosity and SFR by (Rosario et al., 2012). They discovered characteristic trends linking the mean far infrared (FIR) luminosity (L60) and accretion luminosity of AGNs, which are influenced by both luminosity of AGN and redshift. There is no correlation between accretion and SFR at low luminosities, suggesting that these objects are primarily powered by secular processes within their host galaxies. On the other hand, at high AGN luminosities, a notable correlation is evident between L60 and luminosity of AGN, but only for AGN at low and moderate redshifts ($z < 1$).

The importance of the correlation between stellar mass and X-ray luminosity was studied by (Bongiorno et al., 2016) that found that low luminosity AGN may not play a significant role in the process of quenching, especially at higher masses. It is suggested that the feedback from the more powerful AGN (with $logL_{bol} \geq 46[erg/s]$) could be responsible for the suppression of star formation in the host galaxy. Ref (Delvecchio et al., 2015) investigated the connection between stellar mass and X-ray luminosity in a sample of X-ray selected AGN, providing insights into how the stellar mass of the host galaxy influences the AGN's X-ray emission. Also, (Bongiorno et al., 2016) studied the relationship between stellar mass and X-ray luminosity in a large sample of AGNs that span different redshifts, highlighting the evolution of AGN properties with the stellar mass of the host galaxy.

---

*corresponding author
*Email address:* fshaban@sci.cu.edu.eg (F. Shaban )

Several studies have explored the relationship between SFR and X-ray luminosity specifically for radio-loud active galactic nuclei (AGN). Ref (Rosario et al., 2012) found that accretion and SFR are uncorrelated at all redshifts when AGN luminosities are low, which is in line with the hypothesis that secular processes in the host galaxies are the primary source of energy for the majority of low-luminosity AGNs. According to their interpretation, the observations indicate that major mergers are becoming more significant in propelling the expansion of supermassive black holes (SMBHs) and the global star formation in their hosts at high AGN luminosities. Also, they discover evidence that, at high redshifts ($z > 1$), the enhancement of SFR in luminous AGN diminishes or vanishes, indicating that mergers play a less significant role during these periods. On the other hand, (Hardcastle et al., 2009; Mingo, 2014; Drouart et al., 2014) investigated the SFRs in a sample of radio loud AGN, analyzing their X-ray properties and discussing implications for galaxy evolution.
This study investigates correlations between X-ray luminosity ($L_X$), stellar mass ($M_*$), and star formation rate (SFR) for radio-loud (RL) and radio-quiet (RQ) AGN to understand co-evolution and accretion differences. Using XMM-Newton, the DESI AGN Host Galaxy VAC, and CIGALE SED fitting for a uniform sample, we analyze $L_X$-$M_*$-SFR correlations and examine differences across radio classes and accretion modes, addressing limitations of previous studies.

**2. X-ray Observations**

In this section we will present the sample selection method, the reduction and analysis of the sample.

*2.1. Selection Criteria*

The aim of this section is to construct a sample of common objects between both optical and X-ray. We create an X-ray sample by cross-matching the X-ray catalog (4XMM-DR11) with optical quasar catalog (SDSS-DR16) depending on images within 5 arcsecond search radius. For the X-ray catalog (4XMM-DR11), We identified extended objects from point-like sources based on the EPIC extent (< 6 classified as point like source). The total number of point-like sources is 811, out of that, we have 211 source sample. In this section, We follow the selection criteria of Shehata et al. (2021) except for the total number of counts for each source, we choose it > 800 counts that allowed estimating 8904 object.

The final unique X-ray catalog after filtering consists of 6353 point source observed by XMM-NEWTON. The estimated quasars from cross-matching the optical and X-ray catalog are 826 mutual sample with redshift covering range from 0 to 4.964. After reduction we removed sources that have timing mode observations and partial window mode, obtaining 798 quasar. Additionally, We also utilized sdss-dr16, which is a comprehensive catalog of quasars that are highly likely to be optically confirmed quasars, with a 99.8% completeness rate and a contamination rate ranging from 0.3% to 1.3%.

For the optical catalog SDSS-DR16, Quasars were spectroscopically confirmed. The catalog consist of 750,414 quasars with 225,082 newly discovered sources (Lyke et al., 2020). We exclude blazars, galaxies and stars from the optical catalog obtaining 718,850 objects.

*2.2. XMM-NEWTON reduction and analysis*

Here we carried out the analysis covering the energy range $0.5 - 10.0$ keV for PN observations. While all aspects of the data analysis process are automated, it is essential for human intervention to review the X ray images for artifacts and to verify if the source is located in a chip gap, among other issues. We conducted a systematic analysis of the entire spectrum using the zpowerlaw spectral model. It was observed that certain sources benefitted from the inclusion of an additional absorption model, specifically the ZTBABS model, to enhance their results when their reduced_c_stat exceeded 1.5. Moreover, We test if the soft excess is responsible for the great value of reduced_c_stat, we added a blackbody component with a fixed temperature of kT = 0.1 keV for each source using the XSPEC model "zbbody". Adding the blackbody component resulted in a decrease of the C-statistic by at least 2 for some sources.

**3. Radio Classification**

Our sample is constructed through cross matching quasars from SDSS-DR16 with XMM-Newton. To make a detailed study for our sample, we used multi-wavelength analysis. Since quasars' high-energy activities are mainly in radio and X-ray bands, we extend our research and include their radio classification. We matched our 798 sample with the Faint Images of the Radio Sky at Twenty Centimeters (FIRST) catalog (White et al., 1997). Considering a matching radius of 5 arcsec, we got 780 match, which means we have 18 quasars out-off FIRST footprint. Assuming the average power-law slope of $\alpha_v = -0.5$, we calculated the rest-frame 6 cm flux density from the FIRST integrated flux density at 20 cm (White et al., 1997). Ref (Wu and Shen, 2022) fitted the power-law continuum to the spectrum to determine the rest-frame 2500 flux density (Wu and Shen, 2022). Then, estimating the quasar radio loudness parameter ($R$) defining Radio Loud quasars (RLQs) and Radio Quiet Quasars (RQQs) based on the following equation

$$R = \frac{f_{6\,cm}}{f_{2500}} \quad (1)$$

Where the flux densities ($f_v$) at rest-frames 6 cm and 2500 , respectively, are denoted by the expressions $f_{6\,cm}$ and $f_{2500}$ in [mJy]. We used the classification approach by (Miller et al., 2010; Shaban et al., 2022), and classified our sample based on their radio loudness value. For RLQs where $R \geq 10$, we got 118 quasar. For $R < 10$, Radio Intermediate quasars (RIQs), we found only 5 quasars. The rest of the sample is RQQs, which do not have any detectable radio flux, their radio loudness parameter is zero, and they represent the majority of our catalog of 657 quasars. However, studies show that more than 90% of quasars are RQQs and less than 10% are RLQs (Haardt and Maraschi, 1993; Fabian et al., 2015).

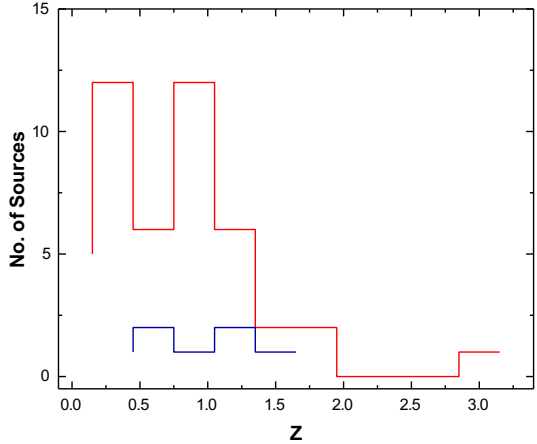

Figure 1: Histogram of redshift distribution for our sample of radio quiet (red) and radio loud (blue) AGN host galaxies.

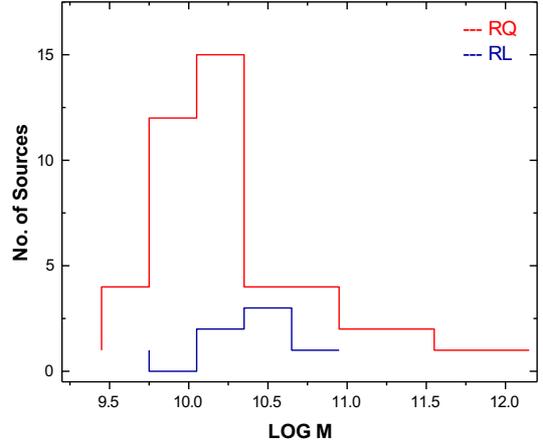

Figure 3: Histogram of stellar mass of host galaxies for our sample of RQ-AGN and RL-AGN

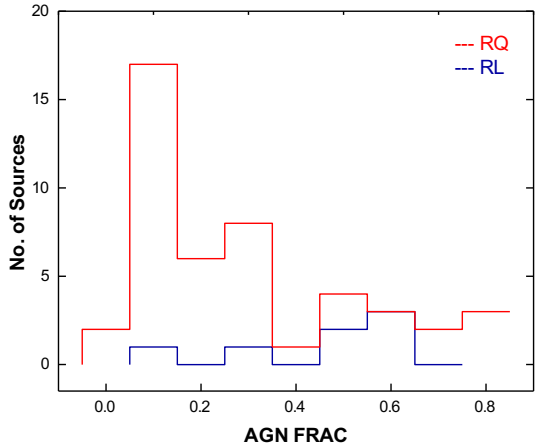

Figure 2: Distributions of the RQ-AGN and RL-AGN fractions for the host galaxies. where 0 means no AGN contribution and 1 means a 100 % AGN contribution.

## 4. The Sample of AGN Host Galaxies

The Dark Energy Spectroscopic Instrument (DESI) (Levi et al., 2013; DESI Collaboration et al., 2016, 2022; Raichoor et al., 2023) survey will measure the distortions of galaxy clustering caused by redshift-space effects, as well as the baryon acoustic feature imprinted on the large-scale structure of the universe, with high precision. The survey aims to accomplish these objectives by performing spectroscopic observations of four different categories of extragalactic sources—nearby bright galaxies (BGS) (Ruiz-Macias et al., 2020; Hahn et al., 2023), luminous red galaxies (LRG) (Zhou et al., 2020, 2023), star-forming emission line galaxies (ELG) (Raichoor et al., 2020, 2023), and quasars (QSO) (Ye`che et al., 2020; Chaussidon et al., 2023).

DESI presents a great chance to study AGNs, because of its large field of view, high spectral resolution, and high data generation. Another advantage of DESI is the high sensitivity, that makes it possible to identify lower-luminosity accretion events. The DESI (https://www.desi.lbl.gov/) project will provide about three million quasar spectra for the next 5 years. With 5000 spectra in a single exposure, DESI is conducting the largest multiobject and high-efficiency spectral survey ever, intending to measure 40 million galaxies and quasars in five years.

The basic sample for this study is a sample of galaxies described in the value added catalog of DESI science collaboration "AGN Host Galaxies Physical Properties VAC" (see https://data.desi.lbl.gov/doc/releases/edr/vac/cigale/). The physical properties of DESI early data release galaxies are included in this value-added catalog (Siudek et al., 2024). These properties are obtained by fitting the spectral energy distribution (SED) with the Code Investigating GALaxy Emission (CIGALE v.22.1; Boquien et al. (2019)), taking into account the contribution of AGN. Approximately 1.3 million galaxies are included in the catalog. Measured parameters include: stellar mass, SFR averaged over 10 Myr, total luminosity of the AGN, fraction of the total IR emission coming from the AGN. rest frame luminosity, and rest frame colors in different bands. The goal of this catalog is to account for a potential AGN contribution while maintaining consistency across the four primary targets (BGS, LRG, ELG, and QSO). By cross matching the above catalogs with DESI AGN Host Galaxies Physical Properties VAC, we confirm the existence of (45 RQ-AGN and 7 RL-AGN), in the redshifts range 0.16 < $z$ < 3.00.

Our radio quiet AGN (RQ-AGN) and radio loud AGN (RL-

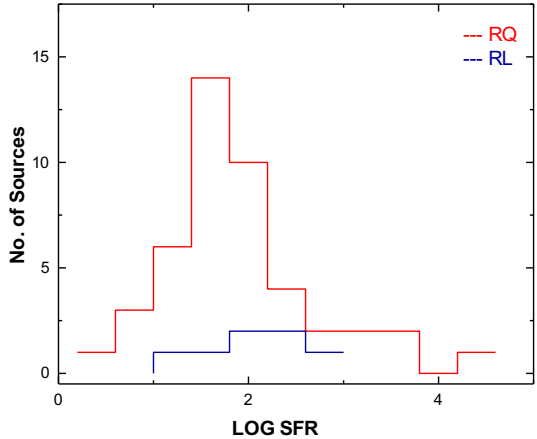

Figure 4: Histogram of SFR for our sample of host galaxies of RQ-AGN and RL-AGN

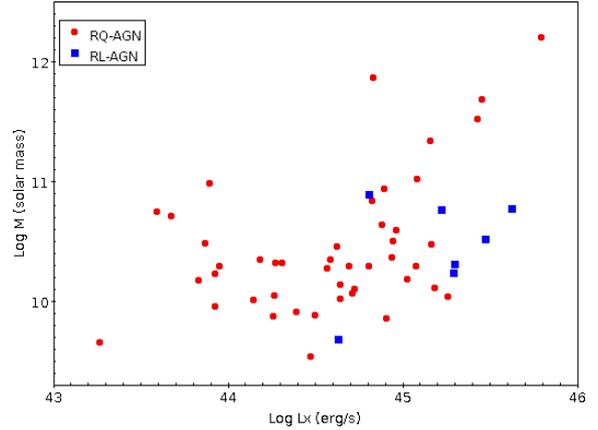

Figure 5: The X-ray luminosity of RQ-AGN and RL-AGN against the stellar mass of their host galaxies

AGN) host galaxies have similar redshift distributions, see Figure 1, with median values of 0.80 and 0.85 respectively. This indicates that our sample is homogeneous, so we could rely on it in further studies. We did not account for RIQ as 5 quasars are not enough to rely on them statistically.

In Figure 2, we show the distribution of RL-AGN and RQ-AGN fraction for the sample of host galaxies. As illustrated in the AGN Host Galaxies Physical Properties VAC, AGN fraction is defined as the fraction of the total infrared (IR) emission coming from the AGN. This plot shows the wide variety of AGN activity found in our sources, such that 0 means no AGN contribution and 1 means a percentage of 100% AGN contribution. It is obvious that our sample of RQ-AGN have a wider range of AGN fractions than the RL-AGN. Also the highest AGN farction is for the radio quiet host galaxies. Our galaxies hosting RQ-AGN show mean value of AGN fraction of 0.32, while the standard deviation value equals 0.24. On the other hand, those hosting RL-AGN have a mean value of 0.50, with standard deviation of 0.20.

Figure 3 shows the distribution of logarithm stellar mass of our sample of AGN host galaxies. We found mean $log(M_s)$ is of about 10.75 and 10.85 for RL-AGN and RQ-AGN, respectively. The standard deviation for $log(M_s)$ is 0.54 for RL-AGN and 0.53 for RQ-AGN. It is clear that the distributions of both samples are quite similar, with RQ-AGN having a slightly higher mean stellar mass than RL-AGN, though the standard deviations indicate a similar spread.

Figure 5 illustrates the correlation between the X-ray luminosity of AGN in erg/s and the stellar mass of the host galaxy in solar masses. It is clear from this figure that RL-AGN (blue symbols) appear at higher stellar masses, potentially indicating that more massive galaxies are more likely to host RL-AGN. On the other hand, RQ-AGN (red symbols) are distributed across the mass range but they may show a different trend with X-ray luminosity. The mean values of $\log L_x$ and $\log M_*$ for RL-AGN are 45.19 and 10.45, respectively. The standard deviation values of $\log L_x$ and $\log M_*$ for RL-AGN are 0.33 and 0.39, respectively. On the other side, the mean values of $\log L_x$ and $\log M_*$ for RQ-AGN equal 44.60 and 10.43, respectively. Also, the standard deviation values of $\log L_x$ and $\log M_*$ for RQ-AGN are 0.54 and 0.56, respectively.

Figure 4 represents the distribution of SFR for RL-AGN and RQ-AGN host galaxies. from this figure, we can say that RL-AGN have a higher average SFR but also a much wider spread in values compared to RQ-AGN, which have a more concentrated distribution. the mean values of $\log SFR$ are 2.12 and 2.24 for RL-AGN and RQ-AGN, respectively. On the other side, standard deviation is 1.25 for RL-AGN and 0.76 for RQ-AGN. This suggests that host galaxies of RL-AGN have a higher average of SFR, but also a much wider spread in values compared to host galaxies of RQ-AGN, which have a more concentrated distribution.

Figure 6 illustrates the correlation between the X-ray luminosity of AGN in erg/s and the SFR of the host galaxy in solar masses/yr. From this figure it is clear that there is a correlation between $\log L_x$ and $\log SFR$ for RL-AGN. The slope of -0.595 in linear regression suggests that star formation does not strongly depend on AGN X-ray luminosity for RL-AGN. This trend indicates that the relationship may be influenced by other factors, such as jet-driven feedback or different galaxy environments. However, RQ-AGN show a moderate positive correlation between AGN X-ray luminosity and SFR of their host galaxies. The positive slope (0.657) using linear regression fitting suggests that as $\log L_x$ increases, $\log SFR$ also tends to increase. This result implies that in RQ-AGN, the AGN activity may be linked to increased star formation, possibly through gas compression mechanisms or a shared gas supply fueling both processes.

Figure 7 illustrates a correlation (for the host galaxies) between the logarithm of stellar mass (log $M$, in solar masses and the logarithm of the star formation rate (log $SFR$). This figure investigates whether there is a systematic difference in SFRs between radio-loud and radio-quiet AGN host galaxies as a function of stellar mass of host galaxy. The correlation coefficient for RL-AGN equals -0.23 (an indication for negative correlation). For RQ-AGN, the correlation coefficient is +0.45, which is an indication for positive correlation. This suggests that RQ-AGN show an increasing trend in SFR with mass for their host galaxies, while RL-AGN show a decreasing trend. In addition to this, the mean and median values for RL-AGN are 3.08 and 3.10, respectively. The mean and median values for RQ-AGN are 2.84 and 2.67, respectively. This indicates that the host galaxies of RL-AGN tend to have slightly higher SFRs on average compared to the host galaxies of RQ-AGN. It is clear from this figure that both populations have a similar spread, which means that the variability in star formation is comparable between RL-AGN and RQ-AGN. It is also obvious that both groups exhibit similar levels of scatter in SFR values.

## 5. Spearman test

It is essential to models of galaxy evolution to comprehend the interaction between the star formation activity within their host galaxies and the growth of supermassive black holes, as indicated by Active Galactic Nucleus (AGN) luminosity. Key parameters in this investigation include the AGN's intrinsic power, often probed by X-ray luminosity ($L_X$), the host galaxy's star formation rate (SFR), and the host's total stellar mass ($M_*$). The relationship between these properties, however, is known to be complex and potentially differs significantly between the major AGN classes: radio-loud (RL) and radio-quiet (RQ) AGN. This section explores the correlations between $L_X$, SFR, and $M_*$ derived from Spearman rank correlation tests performed on distinct samples of radio-loud quasars (RLQ) and radio-quiet (RQ) quasars.

Spearman rank correlation coefficients (ρ) and associated p-values were calculated to assess monotonic relationships between the logarithms of X-ray luminosity (Log$L_X$), star formation rate (LogSFR), and stellar mass (Log$M_*$) for the RLQ and RQ samples separately.

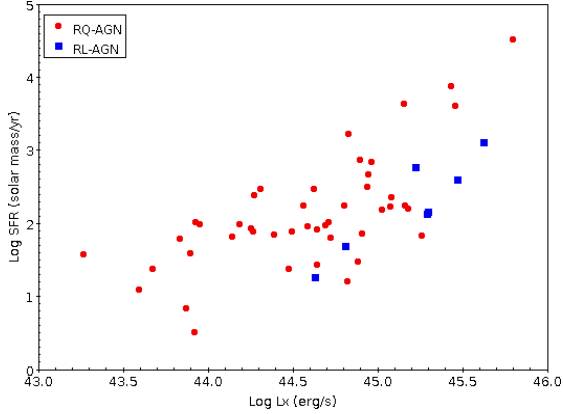

Figure 6: The X-ray luminosity of RQ-AGN and RL-AGN against the SFR of their host galaxies

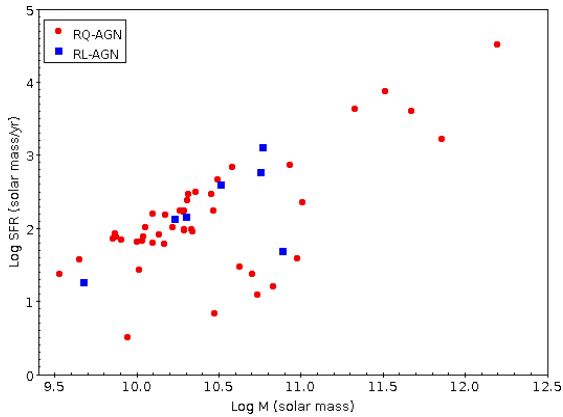

Figure 7: The stellar mass of AGN (radio loud and radio quiet) host galaxies as a function of their SFR

### 5.1. Radio-Quiet Quasars

In the radio-quiet quasar sample, statistically significant positive correlations were observed between all three parameters. A moderate positive correlation was found between Log$L_X$ and LogSFR (ρ = 0.55, $p = 8.2 \times 10^{-5}$), indicating that higher X-ray luminosity is associated with higher star formation rates in these objects. Furthermore, significant, albeit weaker, positive correlations were found between Log$M_*$ and Log$L_X$ (ρ = 0.39, $p = 0.0078$) and between LogSFR and Log$M_*$ (ρ = 0.42, $p = 0.0035$). The correlation between SFR and $M_*$ is consistent with RQ quasar hosts lying on or near the star-forming main sequence (SFMS), where more massive galaxies tend to form stars at a higher rate. The $M_*$-$L_X$ correlation suggests that more massive host galaxies in this RQ sample tend to harbor more luminous AGN. Taken together, these results imply an interconnectedness between host galaxy mass, its ongoing star formation, and the accretion activity onto the central black hole in RQ quasars. This aligns with scenarios where both star formation and AGN activity are fueled by a common gas supply, although the moderate strength of the $L_X$-SFR correlation might reflect complexities introduced by different accretion efficiencies, feedback processes, or the inherent variability timescales of AGN compared to galaxy-wide star formation (e.g., Hickox et al. (2014)).

*5.2. Radio-Loud Quasars*

The radio-loud quasar sample presents a notably different correlation landscape. A strong, statistically significant positive correlation was found between LogL$_X$ and LogSFR ($\rho$ = 0.79, $p$ = 0.036). This suggests a tight coupling between the processes driving X-ray emission and star formation in this particular RLQ sample, potentially even stronger than observed in the RQ sample based on the correlation coefficient. However, in stark contrast to the RQ quasars, no statistically significant correlations were found involving stellar mass. Neither the LogM$_*$-LogL$_X$ relationship ($\rho$ = 0.25, $p$ = 0.59) nor the LogSFR-LogM$_*$ relationship ($\rho$ = 0.46, $p$ = 0.29) showed significant correlation in this RLQ sample.

The lack of significant correlations involving stellar mass in the RLQ sample is intriguing. RL AGN are often found in massive elliptical galaxies (Hardcastle and Croston, 2020), so one might expect correlations if the sample spanned a sufficiently wide range in mass and luminosity. The absence here could suggest that for these RLQs, the instantaneous X-ray luminosity and SFR are not strongly dictated by the total stellar mass, or perhaps that other factors, such as the jet production mechanism or specific feedback processes, play a more dominant role in regulating these properties than the overall host mass. The strong L$_X$-SFR correlation observed is particularly noteworthy, as many studies across the broader AGN population find only weak or flat relationships between instantaneous L$_X$ and SFR, often attributed to AGN variability (e.g., Stanley et al. (2015); Bernhard et al. (2018)). The strong correlation here might hint at specific conditions within this RLQ sample. Possibilities include scenarios where jet production (contributing to radio loudness) is linked to processes that also fuel both accretion (powering L$_X$) and star formation, or perhaps a contribution from the jet itself to the observed X-ray emission which happens to scale with SFR, although the latter requires careful consideration of emission mechanisms.

## 6. Summary and Discussion

In this section, we highlight the significance and innovation of our work, as well as its implications for understanding AGN–host galaxy co-evolution:

- Sample Construction and Methodology: We constructed an X-ray-selected quasar sample by cross-matching the 4XMM-DR11 catalog with the SDSS-DR16 optical catalog within a 5-arcsecond radius. This yielded 826 initial candidates, refined to 798 quasars after excluding sources with poor data quality (e.g., partial window modes, edge artifacts). X-ray spectral analysis was conducted using the zpowerlaw model in the 0.3–10.0 keV band. Sources with reduced c-statistics > 1.5 were re-evaluated with additional absorption models (e.g., ZTBABS or BB) to enhance spectral fits.

- Radio Classification and Multi-wavelength Matching: We matched our refined sample with the FIRST (Faint Images of the Radio Sky at Twenty Centimeters) catalog, identifying 118 Radio-Loud Quasars (RLQs), 5 Radio-Intermediate Quasars (RIQs), and the remainder as Radio-Quiet Quasars (RQQs), based on standard radio loudness parameters. Eighteen quasars were outside the FIRST footprint. RIQs were excluded from further statistical analysis due to small sample size.

- Integration with DESI VAC Data: A key innovation in our study is the integration of the DESI Early Data Release's "AGN Host Galaxies Physical Properties VAC," encompassing 1.3 million galaxies. This catalog provides uniformly derived physical properties (e.g., stellar mass, SFR, AGN IR fraction) using SED fitting with CIGALE v22.1, which includes AGN contributions. We identified 52 AGNs within this VAC sample (45 RQ-AGN and 7 RL-AGN), spanning redshifts $0.16 < z < 3.00$.

- We applied the Spearman rank correlation test to assess the monotonic relationships among total luminosity, star formation rate (log SFR), and stellar mass (log M*) for both types of quasars. Our Spearman rank correlation analysis yielded distinct patterns for radio-loud (RL) and radio-quiet (RQ) AGN populations.

    Analysis of radio-loud AGN (RL-AGN) revealed a strong, significant positive correlation between X-ray luminosity (L$_X$) and star formation rate (SFR) ($\rho$ = 0.79, $p$ = 0.036), suggesting coupled nuclear activity and star formation. However, no significant correlations were found involving stellar mass (M$_*$): L$_X$-M$_*$ ($\rho$ = 0.25, $p$ = 0.589) and SFR-M$_*$ ($\rho$ = 0.46, $p$ = 0.294) were non-significant, indicating a decoupling from host mass in this sample.

    However, radio-quiet AGN (RQ-AGN) showed significant positive correlations across all parameters. Moderate-to-strong correlations were observed for L$_X$-SFR ($\rho$ = 0.55, $p$ = 8.24 $\times$ 10$^{-5}$), L$_X$-M$_*$ ($\rho$ = 0.39, $p$ = 0.0078), and SFR-M$_*$ ($\rho$ = 0.42, $p$ = 0.0035). These results imply an interconnected relationship between AGN luminosity, star formation, and host stellar mass in RQ-AGN, consistent with co-evolution fueled by a common gas supply and adherence to typical star-forming scaling relations.

These results can be contextualized within the broader understanding of AGN populations and their host galaxies. The observed correlations, particularly the differences between RL and RQ AGN, align with and extend previous research.

Regarding AGN accretion physics, our results resonate with studies like Kondapally et al. (2025), which highlighted differences in X-ray characteristics based on accretion modes. They found that RQ-AGN and high-excitation radio galaxies (HERGs), typically associated with radiatively efficient accretion, show higher X-ray luminosities at a given radio power compared to low-excitation radio galaxies (LERGs), which accrete inefficiently. This supports the interpretation that L$_X$ is fundamentally linked to the accretion rate onto the supermassive black hole.

The significant positive $L_X$-SFR correlation observed in both our RL-AGN and RQ-AGN samples suggests a potential link between the mechanisms fueling accretion and star formation. In AGN with efficient accretion (like RQ-AGN and potentially some RL-AGN phases), an abundant supply of cold gas could simultaneously drive high SFR and power luminous X-ray emission. Conversely, the scarcity of cold gas often associated with inefficiently accreting systems (like LERGs) would naturally lead to lower SFR and reduced $L_X$.

Furthermore, our results align with recent investigations into the role of environment. For instance, Mountrichas et al. (2024) found that the relationship between normalized SFR and $L_X$ depends on the large-scale environment. In dense environments, SFR tends to increase with $L_X$, whereas in lower-density fields, the relationship is flatter, with low-to-moderate luminosity AGN showing SFRs comparable to or lower than non-AGN galaxies. This emphasizes that external factors, in addition to intrinsic properties like radio loudness and stellar mass, modulate the co-evolution of AGN and their host galaxies.

In summary, our correlation analysis underscores that the interplay between AGN activity ($L_X$) and host galaxy properties (SFR, $M_*$) is complex and significantly influenced by the AGN type (RL vs. RQ). While RQ-AGN show interconnected scaling between $L_X$, SFR, and $M_*$, consistent with co-evolution driven by gas supply, the RL-AGN in our sample exhibit a strong $L_X$-SFR link but appear decoupled from stellar mass. This highlights the distinct nature of RL systems and the potential influence of jets and feedback mechanisms.

**Software**: Topcat (Taylor, 2017), python packages (Astropy (Astropy Collaboration et al., 2018), Seaborn (Waskom, 2021), Numpy (Harris et al., 2020), and Matplotlib (Hunter, 2007)).

## Acknowledgements

The DESI Legacy Imaging Surveys consist of three individual and complementary projects: the Dark Energy Camera Legacy Survey (DECaLS), the Beijing-Arizona Sky Survey (BASS), and the Mayall z-band Legacy Survey (MzLS). DECaLS, BASS and MzLS together include data obtained, respectively, at the Blanco telescope, Cerro Tololo Inter-American Observatory, NSF's NOIRLab; the Bok telescope, Steward Observatory, University of Arizona; and the Mayall telescope, Kitt Peak National Observatory, NOIRLab. NOIRLab is operated by the Association of Universities for Research in Astronomy (AURA) under a cooperative agreement with the National Science Foundation. Pipeline processing and analyses of the data were supported by NOIRLab and the Lawrence Berkeley National Laboratory (LBNL). Legacy Surveys also uses data products from the Near-Earth Object Wide-field Infrared Survey Explorer (NEOWISE), a project of the Jet Propulsion Laboratory/California Institute of Technology, funded by the National Aeronautics and Space Administration. Legacy Surveys was supported by: the Director, Office of Science, Office of High Energy Physics of the U.S. Department of Energy; the National Energy Research Scientific Computing Center, a DOE Office of Science User Facility; the U.S. National Science Foundation, Division of Astronomical Sciences; the National Astronomical Observatories of China, the Chinese Academy of Sciences and the Chinese National Natural Science Foundation. LBNL is managed by the Regents of the University of California under contract to the U.S. Department of Energy. The complete acknowledgments can be found at https://www.legacysurvey.org/acknowledgment/.